\pdfoutput=1
\documentclass[sigconf]{acmart}
\AtBeginDocument{%
  }

\usepackage{enumitem}
\usepackage{threeparttable}
\usepackage{multirow}
\usepackage{bm}
\usepackage{subfigure}
\usepackage{textcomp}%
\usepackage{changepage}
\usepackage{xcolor}

\usepackage{algpseudocode}
\usepackage{mathtools}

\usepackage{algorithm}
\usepackage{algorithmicx}
\usepackage{amsmath}

\usepackage[group-separator={,}]{siunitx}

\usepackage[justification=justified,skip=3pt]{caption}

\copyrightyear{2025}
\acmYear{2025}
\setcopyright{cc}
\setcctype{by}
\acmConference[XXXXX]{XXXX}{XXXX}{XXXX}
\acmBooktitle{XXXXX}
\acmDOI{XXXXX}
\acmISBN{XXXXX}

\begin{document}

\title{AsarRec: Adaptive Sequential Augmentation for Robust Self-supervised Sequential Recommendation}

\author{Kaike Zhang}
\affiliation{%
  \institution{State Key Laboratory of AI Safety, Institute of Computing Technology, Chinese Academy of Sciences}
  \country{ }
}
\affiliation{%
  \institution{University of Chinese Academy}
  \country{of Sciences, Beijing, China}
}
\email{zhangkaike21s@ict.ac.cn}

\author{Qi Cao}
\affiliation{%
  \institution{State Key Laboratory of AI Safety, Institute of Computing Technology, Chinese Academy of Sciences,}
  \country{Beijing, China}
}
\email{caoqi@ict.ac.cn}
\authornote{Corresponding author}

\author{Fei Sun}
\affiliation{%
  \institution{State Key Laboratory of AI Safety, Institute of Computing Technology, Chinese Academy of Sciences,}
  \country{Beijing, China}
}
\email{sunfei@ict.ac.cn}

\author{Xinran Liu}
\affiliation{%
  \institution{State Key Laboratory of AI Safety, Institute of Computing Technology, Chinese Academy of Sciences,}
  \country{Beijing, China}
}
\email{liuxinran@ict.ac.cn}

\author{Huawei Shen}
\affiliation{%
  \institution{State Key Laboratory of AI Safety, Institute of Computing Technology, Chinese Academy of Sciences,}
  \country{Beijing, China}
}
\email{shenhuawei@ict.ac.cn}

\author{Xueqi Cheng}
\affiliation{%
  \institution{State Key Laboratory of AI Safety, Institute of Computing Technology, Chinese Academy of Sciences,}
  \country{Beijing, China}
}
\email{cxq@ict.ac.cn}

\renewcommand{\shortauthors}{Kaike Zhang et al.}

\begin{abstract}
% Sequential recommender systems have demonstrated strong capabilities in modeling users' dynamic preferences and capturing item transition patterns. However, 

Real-world user behaviors are often noisy due to factors such as human errors, uncertainty, and behavioral ambiguity, which can lead to degraded recommendation performance. To address this issue, recent approaches widely adopt self-supervised learning (SSL), particularly contrastive learning, by generating perturbed views of user interaction sequences and maximizing their mutual information to improve model robustness. However, these methods heavily rely on their pre-defined static augmentation strategies~(where the augmentation type remains fixed once chosen) to construct augmented views, leading to two critical challenges: (1) the optimal augmentation type can vary significantly across different scenarios; (2) inappropriate augmentations may even degrade recommendation performance, limiting the effectiveness of SSL. To overcome these limitations, we propose an adaptive augmentation framework. We first unify existing basic augmentation operations into a unified formulation via structured transformation matrices.  Building on this formulation, we introduce \textbf{AsarRec} (\textbf{A}daptive \textbf{S}equential \textbf{A}ugmentation for \textbf{R}obust Sequential \textbf{Rec}ommendation). To enable stable end-to-end optimization of discrete and strongly constrained augmentations, AsarRec learns to generate transformation matrices by encoding user sequences into probabilistic transition matrices and projecting them into hard semi-doubly stochastic matrices via a differentiable Semi-Sinkhorn algorithm. To ensure that the learned augmentations benefit downstream performance, we jointly optimize three objectives: \textit{diversity} (encouraging distinct views), \textit{semantic invariance} (preserving semantic consistency among views), and \textit{informativeness} (identifying augmentations most beneficial to recommendation). Extensive experiments on four benchmarks under varying noise levels validate the effectiveness of AsarRec, demonstrating its superior robustness and consistent improvements.

% Building on this, we introduce \textbf{AsarRec} (\textbf{A}daptive \textbf{S}equential \textbf{A}ugmentation for \textbf{R}obust Sequential \textbf{Rec}ommendation), which learns to generate transformation matrices by encoding user sequences into probabilistic transition matrices and projecting them into hard semi-doubly stochastic matrices via a differentiable Semi-Sinkhorn algorithm. 

\end{abstract}




\maketitle

\section{INTRODUCTION}
Sequential recommender systems have gained significant popularity due to their ability to effectively capture users' dynamic interests and evolving item transition patterns~\cite{kang2018selfattentive, sun2019bert4rec, zhou2022filterenhanced}. Despite their success, these systems inherently depend on user feedback, which frequently contains noise arising from human errors, uncertainty, and ambiguity in user behaviors~\cite{toledo2016fuzzy, zhang2023robust}. This noise adversely biases learned behavioral patterns, significantly undermining the robustness of recommender models and consequently degrading their performance~\cite{zhang2023robust, wu2016collaborative, shen2025rising}.

\begin{figure}
\centering
\includegraphics[width=0.475\textwidth]{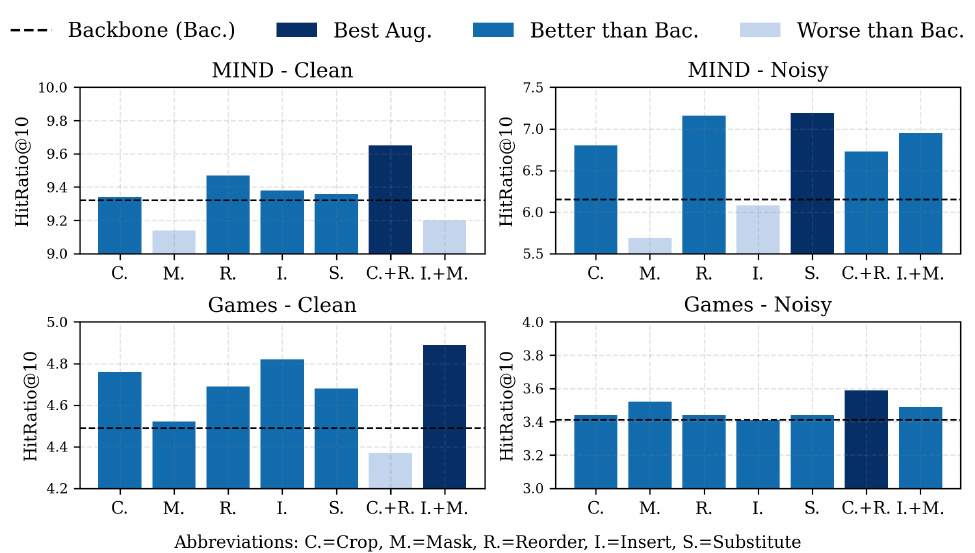}
\caption{Effectiveness of various augmentation methods and their combinations across different data scenarios. In the noisy setting, an additional 20\% random noise is injected into each user's interaction sequence.}
\label{fig:motivation}
\end{figure}

To address these challenges, self-supervised learning (SSL) techniques have emerged as a promising approach for developing robust recommender systems~\cite{zhang2023robust}. These methods typically generate multiple views of a user's interaction sequence by applying data augmentation strategies, such as masking certain interactions or shuffling their order~\cite{67_zhou2020s3, 37_chen2022intent, xie2022contrastive}. The recommender system is then trained to maximize the mutual information between these augmented views~\cite{oord2018representation}, thereby encouraging it to learn noise-invariant representations and improving robustness. Consequently, the performance of SSL methods heavily depends on the design and quality of the augmentation strategies employed. 

However, most existing SSL methods typically adopt \textbf{pre-defined static} augmentation schemes---once an augmentation type is chosen, it is applied uniformly to all users and data scenarios~\cite{67_zhou2020s3, 37_chen2022intent, xie2022contrastive, 23_liu2021contrastive, wang2023knowledge}---rather than adapting the augmentation to dataset characteristics or even to individual users. In practice, we find that \textbf{such static strategies do not consistently yield performance gains and may even degrade the effectiveness of contrastive learning in certain scenarios}, as shown in Figure~\ref{fig:motivation}. For example, in the Games dataset, the Mask operation can bring positive gains by filtering out less informative interactions, whereas in the MIND dataset it instead causes performance degradation, likely due to the removal of semantically important items that disrupt long-range sequential dependencies. Similarly, in the clean setting of Games, the Insert operation, which introduces additional items, effectively enriches the sequence and provides beneficial contextual diversity, leading to substantial improvements. However, under noisy settings---where user histories already contain a high proportion of irrelevant or erroneous interactions---further inserting new items destabilizes training and drastically reduces its effectiveness.

Moreover, certain combinations of augmentations can substantially outperform individual methods (e.g., ``Insert'' combined with ``Mask'' on MIND with 20\% noise), while others can severely underperform, even falling below the results of either augmentation alone (e.g., ``Crop'' combined with ``Reorder'' on clean Games). These observations indicate that the suitability of an augmentation or its combination is highly context-dependent, influenced by dataset characteristics, noise conditions, and user behavior patterns.

The space of possible augmentations and their combinations is vast, yet their effectiveness for SSL varies greatly across settings. Blindly applying mismatched strategies can therefore harm performance, severely constraining the potential of traditional SSL methods. To overcome these limitations, we aim to propose an \textbf{adaptive} augmentation strategy that dynamically tailors augmentation choices to individual users and specific contexts, thereby accommodating diverse scenarios.

However, since different augmentation operations alter interaction sequences in fundamentally different ways, adaptively selecting the most suitable augmentation types and their combinations remains highly challenging. To address this, we first unify classical augmentation methods into a constrained matrix transformation framework. Specifically, given a user interaction sequence $s$, we represent its augmented view $s'$ as the product of the original sequence and a constrained transformation matrix $M$, i.e., $s' = M \cdot s$, where $M$ is defined as a hard semi-doubly stochastic matrix\footnote{A semi-doubly stochastic matrix has each row and column summing to either 0 or 1; a hard semi-doubly stochastic matrix further restricts each element to be exactly 0 or 1}. Building upon this unified formulation, we further propose \textbf{A}daptive \textbf{S}equential \textbf{A}ugmentation for \textbf{R}obust Self-supervised Sequential \textbf{Rec}ommendation, termed \textbf{AsarRec}.

To adaptively generate such transformation matrices for each user, we employ an encoding module. Since directly learning discrete, strongly constrained matrices is challenging due to vanishing or unstable gradients, AsarRec first encodes user sequences into continuous probabilistic transition matrices, and then projects them into hard semi-doubly stochastic matrices using a differentiable semi-Sinkhorn algorithm. Finally, to ensure that the learned augmentations are both effective and reasonable for recommender system, we introduce three optimization constraints: encouraging \textbf{diversity} among transformation matrices to increase the variety of augmentation strategies, preserving \textbf{semantic invariance} by constraining the relative ordering distortion introduced by transformations, and promoting \textbf{informativeness} by constructing challenging augmented views that minimize the mutual information between different representations of the same sequence. Importantly, AsarRec only affects the training phase and does not modify the backbone architecture or inference pipeline. Therefore, it introduces zero inference-time overhead and incurs only marginal additional training cost, making it readily applicable to existing recommender systems.

In this way, AsarRec adaptively identifies suitable and beneficial augmentation strategies (and combinations thereof) for each user sequence and data context, thereby achieving consistently strong performance across diverse recommendation scenarios.

The main contributions of our work are summarized as follows:

\begin{itemize}[leftmargin=*]
    \item We identify key limitations in existing data augmentation strategies and propose to unify sequential data augmentations as a constrained matrix transformation problem, enabling the development of a learnable and adaptive augmentation framework.
    \item Based on this formulation, we present AsarRec, a novel adaptive augmentation framework designed to enhance the robustness and performance of self-supervised sequential recommender systems.
    \item Extensive empirical evaluations on multiple datasets under varying noise levels demonstrate that AsarRec significantly outperforms existing state-of-the-art methods, validating its robustness, adaptability, and effectiveness across diverse settings.
\end{itemize}

\section{RELATED WORK}

\subsection{Sequential Recommender Systems}
Early sequential recommender systems primarily rely on Markov Chain frameworks to capture user interaction patterns~\cite{he2016fusing, he2016vista}. With advancements in neural networks, research evolves toward utilizing architectures such as Recurrent Neural Networks (RNNs)~\cite{hidasi2015session, beutel2018latent} and Convolutional Neural Networks (CNNs)~\cite{tang2018personalized}. The introduction of Transformer-based architectures~\cite{vaswani2017attention}, exemplified by models like SASRec~\cite{kang2018selfattentive} and Bert4Rec~\cite{sun2019bert4rec}, significantly enhances the modeling of dynamic user preferences. More recent work explores alternative approaches, such as Multi-Layer Perceptrons (MLPs)~\cite{zhou2022filterenhanced, zhang2023mlpst}, pushing the boundaries of the field even further. Despite these advancements, the vulnerability of recommender systems to noisy user interactions remains a critical concern, representing a significant challenge in achieving robustness~\cite{zhang2023robust}.

\subsection{Robust Recommender Systems}
Recommender systems built upon user feedback have attracted considerable research attention due to their vulnerability to noise in user interactions~\cite{zhang2023robust, wang2023tutorial, zhang20251st}. Existing strategies to address this challenge can be broadly categorized into two main classes: reweight-based methods and self-supervised learning (SSL) approaches~\cite{zhang2023robust}.

\textbf{Reweight-based Methods.} These methods mitigate the influence of noisy interactions by dynamically adjusting their importance during training~\cite{wang2021denoising, he2024double, lin2023autodenoise, zhang2025personalized}. Specifically, interactions identified as potentially noisy—based on criteria such as high training loss or low confidence—are downweighted or excluded. These methods are model-agnostic and can often be combined with self-supervised learning techniques to further improve robustness.

\textbf{Self-supervised Methods.} In contrast, self-supervised-based approaches enhance robustness by modifying the training objective to make the model inherently resistant to noise. Typically, these methods generate multiple data views by applying various augmentation strategies to user interaction sequences. The model is then trained to maximize the mutual information between augmented views, encouraging the learning of noise-invariant representations~\cite{ma2024madm, wang2022learning, fan2023graph, quan2023robust, zhu2023knowledge, 67_zhou2020s3, 37_chen2022intent, xie2022contrastive, 23_liu2021contrastive, wang2023knowledge}. The commonly used augmentation techniques can be grouped into five types: \textit{Masking} (removing some interactions)~\cite{quan2023robust, 67_zhou2020s3}, \textit{Reordering} (shuffling interaction sequences)~\cite{xie2022contrastive, 23_liu2021contrastive}, \textit{Cropping} (truncating interaction sequences), \textit{Inserting} (adding new interactions), and \textit{Substituting} (replacing interactions)~\cite{xie2022contrastive}. In addition, recent works incorporate informative signals---such as knowledge graphs or item similarity---to generate more meaningful augmentation~\cite{wang2023knowledge, zhu2023knowledge}. However, the optimal augmentation strategy or its combination often varies significantly across datasets and noise conditions. Blindly applying augmentations may lead to suboptimal or even degraded performance, underscoring the need for adaptive augmentation approaches tailored to different scenarios.

\section{PRELIMINARY}
\subsection{Sequential Recommendation}
This section mathematically formulates the task of sequential recommendation. We define the set of users as $\mathcal{U}$ and the set of items as $\mathcal{V}$. Each user $u \in \mathcal{U}$ is associated with a historical interaction sequence $s_u = [v_{1}, v_{2}, \dots, v_{|s_u|}]$, where $v_t \in \mathcal{V}$ denotes an item interacted with by user $u$, and $|s_u|$ represents the length of the sequence. The objective of sequential recommendation is to predict the next item that user $u$ will interact with. This predictive task can be formulated as:
\begin{equation}
v_{|s_u|+1} = \arg\max_{v \in \mathcal{V}} \mathbb{P}(v|s_u),
\end{equation}
where $\mathbb{P}(v|s_u)$ is the conditional probability of item $v$ being the next item following the historical interaction sequence $s_u$.

\begin{figure*}
    \centering 
    \includegraphics[width=6.5in]{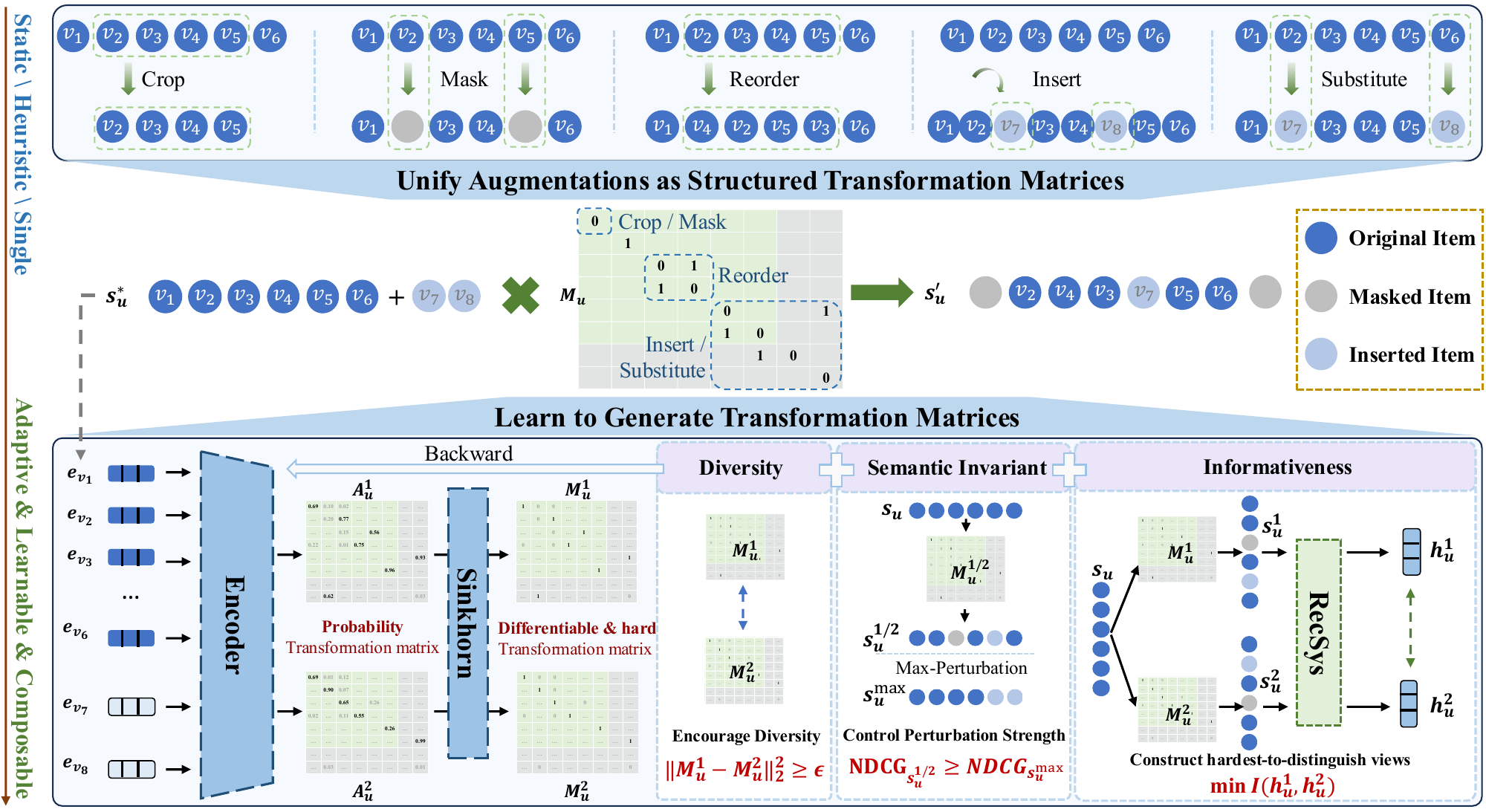}
    \caption{An overview of our proposed framework. The top illustrates five commonly used heuristic augmentation strategies. The middle part shows how we unify these operations as structured transformation matrices. The bottom demonstrates how our model learns to generate effective transformation matrices through a differentiable Sinkhorn-based process, guided by three key objectives: diversity, semantic invariance, and informativeness. Our method enables a transition from static, heuristic, or single-type augmentations to adaptive, learnable, and composable augmentation strategies.}
    \label{fig:framework}
\end{figure*}

\subsection{Self-supervised Learning in Sequential Recommendation}
\label{sec:arug}
Self-supervised learning (SSL) has emerged as a powerful paradigm for enhancing the robustness of sequential recommender systems. The core idea is to construct multiple views of the same user interaction sequence via data augmentation and train the model to maximize their semantic consistency.

Formally, given a user's historical interaction sequence $s_u = [v_{1}, v_{2}, \dots, v_{|s_u|}]$, two augmented views $s_u^1$ and $s_u^2$ can be generated using augmentation strategies. The model is then optimized to maximize the mutual information between the representations of these two views. A general SSL objective can be written as:
\begin{equation}
\mathcal{L}_{\text{SSL}} = - \mathbb{E}_{u \in \mathcal{U}} \left[ I(s_u^1, s_u^2) \right],
\end{equation}
where $I(\cdot, \cdot)$ denotes the mutual information between views. In practice, contrastive loss functions such as InfoNCE~\cite{oord2018representation} or cross-entropy are commonly used to optimize this objective.

A critical component in this framework is the choice of augmentation methods, which determine how diverse and informative the generated views are. Mainstream augmentation strategies for sequential data can be grouped into five basic types:
\begin{itemize}[leftmargin=*]
\item \textbf{Crop:} Truncating the user interaction history by retaining a continuous subsequence.
\item \textbf{Mask:} Discarding a portion of interactions in the sequence.
\item \textbf{Reorder:} Shuffling the order of items within the sequence.
\item \textbf{Insert:} Inserting randomly selected interactions into the sequence.
\item \textbf{Substitute:} Replacing existing interactions with sampled alternatives.
\end{itemize}
Beyond these basic types, recent studies have explored more informative augmentation strategies by incorporating external signals, such as item similarity or knowledge graphs. These methods typically extend or combine the above five operations to produce semantically meaningful perturbations.

\section{METHOD}
\label{sec:method}
In this section, we introduce \textbf{AsarRec} (an Adaptive Sequential Augmentation for Robust Self-supervised Sequential Recommender System), designed to enhance self-supervised sequential recommendation. Unlike existing contrastive learning methods that rely on static, heuristics strategies, AsarRec dynamically learns suitable augmentation---and their combinations---in a data-driven manner for different users.

\subsection{Unified Matrix-based Formulation for Sequential Augmentation}
Since different augmentation operations modify interaction sequences in fundamentally different ways, selecting a suitable strategy adaptively is challenging. As a foundational step toward this goal, we propose a unified matrix-based formalization for existing sequential data augmentation methods.

\textbf{Unified Operations for Crop, Mask, and Reorder:}
For a given user interaction sequence $s$, we represent its augmented version $s'$ through a constrained binary matrix transformation $M_o \in \{0,1\}^{|s| \times |s|}$, as illustrated in middle part of Figure~\ref{fig:framework}:
\begin{equation}
s' = M_o \cdot s,
\end{equation}
where $M_o$ is a hard semi-doubly stochastic matrix satisfying:
\begin{equation}
\sum M_{o,i,*} \in \{0,1\}, \quad \sum M_{o,*,j} \in \{0,1\}.
\label{eq:semi}
\end{equation}
Here, $\sum M_{o,i,*} = 0$ indicates that the item $v_i$ is removed from the sequence, corresponding to Mask (or Crop, which can be viewed as masking consecutive items). $M_{o,i,j} = 1$ means that the item $v_i$ from the original sequence $s$ is placed at position $j$, corresponding to the Reorder operation.

\textbf{Unified Operations for Insert and Substitute:}
Since insert and substitute operations introduce new items, we unify these operations by applying random padding. Specifically, given the original sequence $s$, we randomly pad $k$ new items not present in $s$, resulting in an expanded sequence:
\begin{equation}
s^* = [v_1, v_2, \dots, v_{|s|}, v_{|s|+1}, \dots, v_{|s|+k}].
\end{equation}

We then extend the transformation matrix $M_o$ into a new hard semi-doubly stochastic matrix $M \in \{0,1\}^{(|s| + k) \times (|s|+k)}$, subject to the same constraints as in Formula~\ref{eq:semi}, which acts upon the expanded sequence $s^*$. If $\sum\nolimits_{i \ge |s|, j < |s|} M_{i,j} \ge 1,$ it indicates that newly introduced items are inserted or substituted into the user's historical interaction sequence.

Through this unified formulation, the five typical sequential augmentation operations---Reorder, Mask, Crop, Insert, and Substitute---are all represented within a structured, hard semi-doubly stochastic matrix transformation framework, enabling a consistent and learnable representation of diverse augmentation strategies.

\subsection{Generation of Sequential Transformation Matrices}

Having established a unified matrix transformation for sequential augmentation, our next goal is to learn suitable hard semi-doubly stochastic matrices $M$ for each user's interaction sequence. Given a user's interaction sequence $s_u^*$ (including padding), we aim to generate two different corresponding transformation matrices $M_u^1$ and $M_u^2$ for two different views, as shown in bottom of Figure~\ref{fig:framework}.

Since $M_u^1$ and $M_u^2$ are strongly constrained discrete matrices, directly learning them is challenging due to vanishing or unstable gradients. To mitigate this, we introduce an encoding module that first maps a user's historical interaction sequence into a continuous probabilistic transition matrix. Specifically, we represent the padded user sequence $s_u^*$ as an embedding sequence $\bm{s}_u^* \in \mathbb{R}^{|s_u^*| \times d}$ using the current recommender system, where $d$ is the embedding dimension. These embeddings are then projected into a $d'$-dimensional space via a shared linear transformation $W \in \mathbb{R}^{d \times d'}$, and subsequently fed into two independent self-attention modules to generate probabilistic projection matrices $A_u^{(z)} \in \mathbb{R}^{|s_u^*| \times |s_u^*|}$ for $z \in \{1,2\}$:
\begin{equation}
A_u^{(z)} = \mathrm{softmax}\Bigg(
\frac{ (\bm{s}_u^* W^\top) W_Q^{(z)} \Big( (\bm{s}_u^* W^\top) W_K^{(z)} \Big)^\top }{\sqrt{d'}}
\Bigg),
\quad z \in \{1,2\},
\end{equation}
where $W_Q^{(z)}, W_K^{(z)} \in \mathbb{R}^{d' \times d'}$ are learnable parameters, and $A_{u,i,j}$ denotes the probability of mapping item $v_i$ in $s_u^*$ to position $j$.

To obtain the final hard semi-doubly stochastic matrices $M_u^{(z)} \in \{0,1\}^{|s_u^*| \times |s_u^*|}$ while preserving \textbf{differentiability}, we apply the Semi-Stochastic-Sinkhorn algorithm to $A_u^{(z)}$:
\begin{equation}
M_u^{(z)} = \mathrm{Semi\text{-}Sinkhorn}(A_u^{(z)}), \quad z \in \{1, 2\},
\end{equation}
where $\mathrm{Semi\text{-}Sinkhorn}(\cdot)$ iteratively normalizes rows and columns to satisfy approximate semi-doubly stochastic constraints. Compared with softmax-based projections, this procedure mitigates gradient scaling imbalance or distortion, ensuring stability even after multiple iterations while progressively approximating a hard binary structure. 
% Finally, we adopt the Gumbel-Sinkhorn trick to approximate the discrete sampling step in a differentiable manner. 
\textbf{The complete procedure is detailed in Algorithm~\ref{alg:semi}}.

Through this differentiable sampling process, we obtain the final hard semi-doubly stochastic matrices $M_u^1$ and $M_u^2$, which serve as the foundation for subsequent adaptive augmentation learning.

\subsection{Learning Objectives for Transformation Matrices}
To ensure that the learned transformation matrices are both reasonable and beneficial to downstream recommendation tasks, we introduce three constraints to jointly optimize the encoding model. Each objective in AsarRec is designed to address a specific failure mode in self-supervised sequential recommendation.
Specifically, (1) the diversity constraint prevents mode collapse where identical augmentations trivially minimize the loss;
(2) the semantic invariance constraint avoids semantically invalid or destructive perturbations;
and (3) the informativeness constraint encourages non-trivial but beneficial augmentations that improve robustness.

\textbf{Diversity Constraint:}
We expect the two learned transformation matrices $M_u^1$ and $M_u^2$ for the same user interaction sequence $s_u^*$ to differ from each other, avoiding redundant augmentation patterns. To achieve this, we introduce a diversity constraint that enforces a minimum difference between them under the $l_2$ norm. The corresponding loss is defined as:
\begin{equation}
\mathcal{L}_{\mathrm{div}} =  \mathbb{E}_{u \in \mathcal{U}} \left[  \max\left( 0, \epsilon - | M_u^1 - M_u^2 |_2^2 \right)\right],
\end{equation}
where $\epsilon$ is a predefined threshold.

\textbf{Semantic Invariance Constraint.}
\label{sec:semantic}
To prevent excessive perturbations from fundamentally altering the semantics of the original sequence, we introduce a constraint based on \textbf{Sequence-aware NDCG}\footnote{Unlike the conventional NDCG that computes relevance solely based on predefined relevance labels, Sequence-aware NDCG assigns relevance scores according to the temporal positions of items in the original sequence, thereby emphasizing the preservation of recency-aware semantics in sequential recommendation.}, which measures the ranking consistency between the original and transformed sequences while accounting for the sequential order of interactions. Given a user interaction sequence $s_u = [v_1, v_2, \dots, v_{|s_u|}]$, we assign a relevance score $r_i$ to each item $v_i$ based on its position in the original sequence:
\begin{equation}
r_i = \text{position}(v_i),
\end{equation}
where $\text{position}(v_i)$ denotes the index of $v_i$ in $s_u$. The most recent item (i.e., the one at the end of the sequence) receives the highest relevance score. For instance, in a sequence of length 50, the relevance scores range from 1 (oldest) to 50 (most recent).

To construct the reference for worst-case semantic degradation, we simulate the maximum perturbation by replacing the recent $\lfloor \gamma |s_u| \rfloor$ items---those with highest relevance---with irrelevant items (i.e., with relevance scores set to 0). Let $s^{\text{max}}_u$ denote the resulting sequence, and $\text{NDCG}_{s^{\text{max}}_u}$ its Sequence-aware NDCG score.

Given two augmented sequences $s_u^1$ and $s_u^2$, we compute their respective Sequence-aware NDCG scores $\text{NDCG}_{s_u^1}$ and $\text{NDCG}_{s_u^2}$ based on the resulting ranking after transformation. The semantic invariance loss is then defined as:
\begin{equation}
\mathcal{L}_{\text{NDCG}} = \mathbb{E}_{u \in \mathcal{U}} \left[ \sum_{z=1}^2 \max(0, \text{NDCG}_{s^{\text{max}}_u} - \text{NDCG}_{s_u^z}) \right].
\end{equation}

This constraint encourages the learned transformation matrices to preserve the relative ranking of semantically important items, ensuring that the perturbed views remain faithful to the original user intent while retaining semantic plausibility.

\textbf{Informativeness Constraint:}
To ensure that the learned augmentations are informative, i.e., the downstream recommendation model can benefit from the contrast between the two augmented views, we aim to minimize the mutual information between them. This encourages the generation of challenging augmentation pairs that introduce meaningful challenges to the model, thereby offering greater potential gains in noise-invariant representations learning. To achieve this, we adopt the InfoNCE loss to constrain the mutual information between the two augmented sequences after encoding by the recommendation model:
\begin{equation}
\mathcal{L}_{\mathrm{info}} = \mathbb{E}_{u \in \mathcal{U}} \left[ \log \frac{
\exp( \mathrm{sim}(f(\bm{s}_u^{1}), f(\bm{s}_u^{2})) / \tau )
}{
\sum \exp( \mathrm{sim}(f(\bm{s}_u^{1}), f(\bm{s}_j^{2})) / \tau )
} \right],
\end{equation}
where $f(\cdot)$ denotes the recommendation model, $\mathrm{sim}(\cdot)$ is the cosine similarity function, $\tau$ is the temperature coefficient, and the denominator sums over $N$ negative samples in the batch.

\begin{algorithm}[t]
\caption{Differentiable Semi-Stochastic Sinkhorn Sampling}
\begin{algorithmic}[1]
\Require Probability matrix $A$, threshold $\delta$, iteration count $T$
\State $S \gets A$
\For{$t = 1$ to $T$}
    \State Row normalization: $S \gets S / S.\text{row\_sum}$
    \State Column normalization: $S \gets S / S.\text{col\_sum}$
    \For{each row or column}
        \If{difference between max and min $< \delta$}
            \State Set this row or column to zero
        \EndIf
    \EndFor
\EndFor
\State $M_{\text{hard}} \gets \mathrm{argmax\_per\_row\_col}(S)$
\State $M \gets (M_{\text{hard}} - S).\text{detach}() + S$
\State \Return $M$
\end{algorithmic}
\label{alg:semi}
\end{algorithm}

\textbf{Discussion.} 
It is worth noting that the \textbf{Diversity} and \textbf{Informativeness} constraints exhibit partially overlapping objectives, as both encourage the generation of distinct augmented views. However, their emphases differ: the Diversity constraint focuses on maximizing augmentation variation and fully utilizing the perturbation budget, operating directly on the transformation matrices $M_u^1$ and $M_u^2$, whereas the Informativeness constraint aims to identify and perturb highly informative behavioral patterns, operating on the representations produced by the recommender system. In this context, situations may arise where the learned representations differ substantially while $M_u^1$ and $M_u^2$ remain similar, or vice versa, resulting in augmentations that fail to meet the intended objectives. 

Moreover, our empirical observations show that without the Diversity constraint, the two augmented views can degenerate into identical sequences. 
When such trivial pairs are passed into the InfoNCE objective, they may trigger gradient explosion or training instability. 
Thus, the Diversity constraint not only promotes augmentation diversity and effective perturbation budget utilization, but also plays a crucial role in ensuring training stability.

Integrating the three constraints, our final joint loss function is defined as:
\begin{equation}
\mathcal{L} = \mathcal{L}_{\mathrm{info}} + \lambda_{\mathrm{div}} \mathcal{L}_{\mathrm{div}} + \lambda_{\mathrm{NDCG}} \mathcal{L}_{\mathrm{NDCG}},
\end{equation}
where $\lambda_{\mathrm{div}}, \lambda_{\mathrm{NDCG}}$ are hyperparameters that control the contribution of each constraint to the overall objective.

% \subsection{Unified Matrix Transformation for Sequential Augmentation}
% \input{Section/4-Subsection/1-Unify}

% \subsection{How to Generate Sequential Transformation Matrix}
% \input{Section/4-Subsection/2-Learn2Aug}

% \subsection{How to Learn Sequential Transformation Matrix}
% \input{Section/4-Subsection/3-Train}

\section{EXPERIMENTS}
% \begin{table}[t]
%   \centering
%     \caption{Dataset statistics}
%     \resizebox{0.47\textwidth}{!}{
%         \begin{threeparttable}

% \begin{tabular}{lrrrrr}
%     \toprule
%     \textbf{ DATASET } & \textbf{ \#Users } & \textbf{ \#Items } & \textbf{\#Ratings}  & \textbf{Avg.length} & \textbf{Sparsity}\\
%     \midrule
%      Games  & 61,521 & 33,243 & 541,789 & 8.8 & 99.97\% \\
%      Arts  & 71,364 & 61,505 & 600,989 & 8.4 & 99.99\% \\ 
%      MIND  & 152,909 & 63,608 & 4,186,679 & 27.4 & 99.96\% \\ 
%     \bottomrule
%     \end{tabular}
%         \end{threeparttable}
%     }
%   \label{tab:datasets}%
% \end{table}%

\begin{table}[t]
  \centering
  \caption{Dataset statistics}
  \resizebox{0.47\textwidth}{!}{
    \begin{threeparttable}
      \begin{tabular}{lrrrrr}
        \toprule
        \textbf{Dataset} & \textbf{\#Users} & \textbf{\#Items} & \textbf{\#Interactions} & \textbf{Avg. length} & \textbf{Sparsity} \\
        \midrule
        Games  & 61,521 & 33,243 & 541,789 & 8.8 & 99.97\% \\
        Arts   & 71,364 & 61,505 & 600,989 & 8.4 & 99.99\% \\
        MIND   & 152,909 & 63,608 & 4,186,679 & 27.4 & 99.96\% \\
        ML-20M & 138,493 & 27,278 & 20,000,263 & 144.4 & 99.47\% \\
        \bottomrule
      \end{tabular}
    \end{threeparttable}
  }
  \label{tab:datasets}
\end{table}

In this section, we conduct extensive experiments to answer the following research questions (\textbf{RQs}):
\begin{itemize}[leftmargin=*]
    \item \textbf{RQ1:} How does \textbf{AsarRec} perform compared to state-of-the-art augmentation methods under both clean and noisy conditions?
    \item \textbf{RQ2:} How well does \textbf{AsarRec} generalize across different backbone models, which types or combinations of augmentations does it tend to prefer, and what is its computational complexity?
    \item \textbf{RQ3:} How sensitive is \textbf{AsarRec} to key hyperparameter variations, and how does each component contribute to its overall effectiveness (ablation study)?
\end{itemize}

\begin{table*}[t]
    \centering
    \caption{Recommendation performance of different argumentation methods. The highest scores are in bold, and the runner-ups are with underlines. A significant improvement over the runner-up is marked with * (i.e., two-sided t-test with $0.05 \le p < 0.1$) and ** (i.e., two-sided t-test with $p < 0.05$). Results that lead to performance degradation are shaded in gray.}
    \resizebox{\textwidth}{!}{

\begin{tabular}{lcccccccccccc}
    \toprule
    \multicolumn{1}{c}{\multirow{3}{*}{\textbf{Model}}} & \multicolumn{4}{c}{\textbf{Games} } & \multicolumn{4}{c}{\textbf{Arts} } & \multicolumn{4}{c}{\textbf{MIND} }  \\
    \cmidrule(lr){2-5} \cmidrule(lr){6-9} \cmidrule(lr){10-13}
    & \multicolumn{2}{c}{\textbf{HR} } & \multicolumn{2}{c}{\textbf{NDCG} } & \multicolumn{2}{c}{\textbf{HR} } & \multicolumn{2}{c}{\textbf{NDCG} } & \multicolumn{2}{c}{\textbf{HR} } & \multicolumn{2}{c}{\textbf{NDCG} } \\
    \cmidrule(lr){2-3} \cmidrule(lr){4-5} \cmidrule(lr){6-7} \cmidrule(lr){8-9} \cmidrule(lr){10-11} \cmidrule(lr){12-13}
    & \textbf{@10} & \textbf{@20} & \textbf{@10} & \textbf{@20} & \textbf{@10} & \textbf{@20} & \textbf{@10} & \textbf{@20} & \textbf{@10} & \textbf{@20} & \textbf{@10} & \textbf{@20} \\
    
    \midrule
    \textbf{SASrec}& 0.0449 & 0.0745 & 0.0231 & 0.0311 & 0.0367 & 0.0528 & 0.0214 & 0.0252 & 0.0932 & 0.1500 & 0.0471 & 0.0613 \\
~+\textbf{Crop}& 0.0476 & 0.0777 & 0.0244 & 0.0319 & 0.0394 & 0.0547 & 0.0225 & 0.0264 & 0.0934 & 0.1500 & \textcolor{gray}{0.0469} & \textcolor{gray}{0.0611} \\
~+\textbf{Mask}& 0.0452 & \textcolor{gray}{0.0738} & \textcolor{gray}{0.0228} & \textcolor{gray}{0.0300} & 0.0375 & \textcolor{gray}{0.0525} & 0.0230 & 0.0268 & \textcolor{gray}{0.0914} & \textcolor{gray}{0.1467} & \textcolor{gray}{0.0459} & \textcolor{gray}{0.0597} \\
~+\textbf{Reorder}& 0.0469 & 0.0766 & 0.0238 & 0.0313 & 0.0383 & 0.0540 & 0.0217 & 0.0256 & \underline{0.0947} & \underline{0.1515} & 0.0475 & \underline{0.0618} \\

~+\textbf{Insert}& \underline{0.0482} & 0.0780 & 0.0243 & 0.0318 & 0.0394 & \underline{0.0550} & 0.0230 & 0.0269 & 0.0938 & \textcolor{gray}{0.1491} & 0.0475 & 0.0614 \\
~+\textbf{Sub}& 0.0468 & 0.0765 & 0.0239 & 0.0313 & \textcolor{gray}{0.0362} & \textcolor{gray}{0.0527} & \underline{0.0234} & \underline{0.0275} & 0.0936 & 0.1501 & \textcolor{gray}{0.0467} & \textcolor{gray}{0.0609} \\
\cmidrule{2-13}
~+\textbf{CoSeRec}$_{\text{Insert}}$& 0.0480 & 0.0780 & \underline{0.0246} & 0.0321 & 0.0384 & 0.0539 & 0.0223 & 0.0262 & 0.0945 & 0.1507 & \underline{0.0476} & 0.0617 \\
~+\textbf{CoSeRec}$_{\text{Sub}}$& 0.0477 & \underline{0.0782} & 0.0246 & \underline{0.0322} & 0.0383 & 0.0538 & 0.0222 & 0.0260 & 0.0938 & 0.1503 & 0.0471 & 0.0613 \\
~+\textbf{CL4Rec}& 0.0479 & 0.0776 & 0.0244 & 0.0318 & \underline{0.0394} & 0.0544 & 0.0231 & 0.0269 & 0.0939 & 0.1507 & 0.0472 & 0.0616 \\
~+\textbf{S3Rec}& 0.0475 & 0.0770 & 0.0244 & 0.0318 & 0.0386 & 0.0536 & 0.0232 & 0.0269 & \textcolor{gray}{0.0928} & 0.1506 & \textcolor{gray}{0.0465} & \textcolor{gray}{0.0610} \\
\cmidrule{2-13}
~+\textbf{AsarRec}&\textbf{0.0512**} &\textbf{0.0837**} &\textbf{0.0256} &\textbf{0.0337*} &\textbf{0.0429*} &\textbf{0.0616**} &\textbf{0.0265**} &\textbf{0.0312*} &\textbf{0.1030**} &\textbf{0.1658**} &\textbf{0.0511**} &\textbf{0.0669**} \\
\multicolumn{1}{c}{Gain$^1$}& +6.40\% $\uparrow$& +7.08\% $\uparrow$& +4.07\% $\uparrow$& +4.71\% $\uparrow$& +8.95\% $\uparrow$& +12.10\% $\uparrow$& +13.39\% $\uparrow$& +13.49\% $\uparrow$& +8.83\% $\uparrow$& +9.47\% $\uparrow$& +7.39\% $\uparrow$& +8.32\% $\uparrow$\\
    
    \bottomrule
\end{tabular}
    }
    \begin{flushleft}
    \begin{itemize}
    \footnotesize 
         \item[1] The relative percentage increase of AsarRec's metrics to the best value of other baselines' metrics, i.e., $\left(\min\left(\mathrm{HR}_\mathrm{AsarRec} - \mathrm{HR}_\mathrm{Beslines}\right)\right)/ \min(\mathrm{HR}_\mathrm{Beslines})$.
    \end{itemize}
    \end{flushleft}
\label{tab:performance}%
\end{table*}

\subsection{Experimental Setup}
\label{sec:exp_setup}

\subsubsection{Datasets}
In our evaluation of AsarRec, we employ three widely recognized datasets: the Amazon review datasets (\textbf{Games} and \textbf{Arts})~\cite{ni2019justifying}, the \textbf{MIND} news recommendation dataset~\cite{wu2020mind}, and the large-scale \textbf{MovieLens-20M (ML-20M)} dataset. For the Amazon and ML-20M datasets, all users are considered. For the MIND dataset, a subset of users is sampled following~\cite{li2023exploring}. Consistent with existing practices~\cite{rendle2012bpr, kang2018selfattentive, zhang2024lorec}, we exclude users with fewer than 5 interactions. For each user, the data split process involves (1) using the most recent action for testing, (2) the second most recent action for validation, and (3) all preceding actions (up to 50) for training~\cite{kang2018selfattentive, li2023exploring, yuan2023where}. Dataset statistics are provided in Table~\ref{tab:datasets}.

\subsubsection{Backbone Models}
We employ three backbone sequential recommender systems:
\begin{itemize}[leftmargin=*]
    \item \textbf{GRU4rec}~\cite{hidasi2015session} utilizes Recurrent Neural Networks~\cite{cho2014learning} to model user interaction sequences in session-based recommendations.
    \item \textbf{SASrec}~\cite{kang2018selfattentive} employs a multi-head self-attention mechanism in Transformer~\cite{vaswani2017attention} for sequential recommendations.
    \item \textbf{FMLPrec}~\cite{zhou2022filterenhanced} is an all-Multilayer Perceptron model with a learnable filter-enhanced block for noise reduction in embeddings for sequential recommendations.
\end{itemize}
\textbf{Due to space limitations}, we predominantly show results using SASrec as the backbone model in Section~\ref{sec:rq1}. Results for GRU4rec and FMLPrec are described in Section~\ref{sec:rq2}.

\subsubsection{Baselines}
We abstract five basic types of augmentation strategies: \textbf{Crop}, \textbf{Mask}, \textbf{Reorder}, \textbf{Insert}, and \textbf{Substitute}, from existing works~\cite{xie2022contrastive, wang2022explanation}, as described in Section~\ref{sec:arug}.  

Beyond these basic operations, some studies enhance augmentation quality by incorporating informative signals such as item similarity~\cite{23_liu2021contrastive}, user metadata~\cite{xiao2024generic}, or their combinations. For fair comparison and due to space constraints, we select the following representative methods:  
\begin{itemize}[leftmargin=*]
    \item \textbf{CoSeRec}~\cite{23_liu2021contrastive}: Inserts or replaces an item in the sequence with the most similar item retrieved from external knowledge sources.
    \item \textbf{CL4Rec}~\cite{xie2022contrastive}: To maintain training stability, we adopt its setting where \textit{Mask} and \textit{Reorder} are applied as two complementary data views for contrastive learning.
    \item \textbf{S3Rec}~\cite{67_zhou2020s3}: For fairness, we employ its \textit{predictive augmentation}, which reconstructs masked segments of the sequence using a self-supervised prediction loss.
\end{itemize}

\subsubsection{Evaluation Metrics}
We adopt standard metrics widely employed in the field. The primary metrics for evaluating recommendation performance are the top-$k$ metrics: Hit Ratio at $K$ ($\mathrm{HR}@K$) and Normalized Discounted Cumulative Gain at $K$ ($\mathrm{NDCG}@K$), as described in~\cite{zhang2023robust, zhang2024lorec, kang2018selfattentive}. Following~\cite{zhang2023robust}, we set $K=10$ and $K=20$.

\subsubsection{Implementation Details} 
For backbone models, the learning rate is selected from \{0.1, 0.01, $\dots, 1 \times 10^{-5}$\}, and the weight decay is chosen from \{0, 0.1, $\dots, 1 \times 10^{-5}$\}. The backbone model architectures follow their original implementations as described in the corresponding publications. For all augmentation methods, we set the perturbation budget $\gamma$ to 10\%. In the case of AsarRec, we set the threshold $\epsilon$ to $20.0$. The training weights $\lambda_{\text{div}}$ is selected from \{1.0, 2.0, 5.0, 50.0, 100.0\}; $\lambda_{\text{NDCG}}$ is selected from \{0.1, 0.5, 1.0, 2.0, 5.0\}. The candidate number $k$ is selected from \{2, 3, 5, 10, 20\}. Our implementation is available at the following link\footnote{\url{https://github.com/Kaike-Zhang/AsarRec}}.
% \footnote{\url{https://anonymous.4open.science/r/AsarRec}}.

\begin{figure*}
    \centering
    \includegraphics[width=0.975\linewidth]{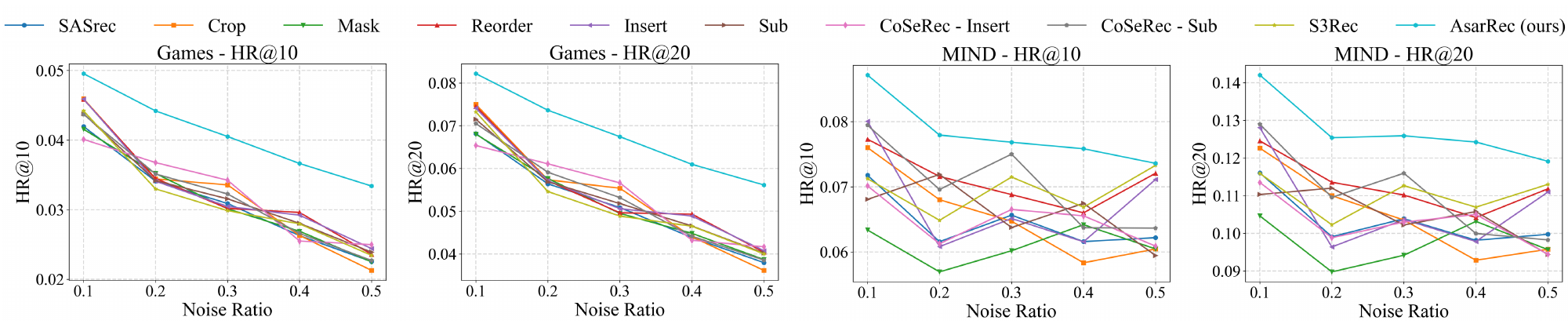}
    \caption{Recommendation performance of different argumentation methods across various noise ratios.}
    \label{fig:all_noise}
\end{figure*}

\subsection{Performance Comparison~(RQ1)}
\label{sec:rq1}
In this section, we address \textbf{RQ1} by focusing on two key aspects: recommendation performance and robustness against noise.

\subsubsection{Recommendation Performance}

We first evaluate the effectiveness of AsarRec across three real-world datasets without injecting additional synthetic noise. Table~\ref{tab:performance} reports HR@10/20 and NDCG@10/20 for each augmentation method. Based on the results, we derive the following observations:

\begin{itemize}[leftmargin=*]
    \item \textbf{AsarRec consistently achieves state-of-the-art performance.} On average, AsarRec achieves relative gains of \textbf{+7.55\%} in HR@10, \textbf{+8.24\%} in HR@20, \textbf{+6.61\%} in NDCG@10, and \textbf{+8.17\%} in NDCG@20 compared to the strongest baseline. These consistent improvements highlight the effectiveness of learning augmentation in a data- and user-adaptive manner.

    \item \textbf{The best augmentation method (excluding AsarRec) varies across datasets and metrics.} No single baseline consistently performs best across all settings. For example, Insert yields the best HR@10 on Games, CoSeRec is best for HR@20 on Games, while Reorder performs best for HR@10 on MIND. This inconsistency reveals the high sensitivity of fixed augmentation strategies to dataset characteristics.

    \item \textbf{Basic augmentation methods provide limited and inconsistent gains.} Insert and Reorder occasionally show competitive results, but their performance fluctuates across datasets. Other methods (e.g., Substitute) often degrade performance.
\end{itemize}

In addition to the above benchmarks, we further evaluate AsarRec on ML-20M. Due to space limitations, we report results by comparing AsarRec with the backbone model and the strongest baseline on this dataset. The results are summarized in Table~\ref{tab:ml20m}. These results demonstrate that AsarRec maintains consistent improvements even at a much larger scale, validating its scalability and effectiveness in large-scale sequential recommendation scenarios.

\begin{table}[t]
    \centering
    \caption{Recommendation performance on the ML-20M.}
    \resizebox{0.42\textwidth}{!}{
        \begin{threeparttable}
        \begin{tabular}{lcccc}
            \toprule
            \textbf{Metric} & \textbf{Base} & \textbf{SOTA} & \textbf{AsarRec} & \textbf{Gain} \\
            \midrule
            HR@10   & 0.2175 & 0.2231 & \textbf{0.2317} & +3.86\% $\uparrow$ \\
            NDCG@10 & 0.1203 & 0.1228 & \textbf{0.1261} & +2.69\% $\uparrow$ \\
            \bottomrule
        \end{tabular}
        \end{threeparttable}
    }
    \label{tab:ml20m}
\end{table}

\subsubsection{Robustness against Noise}

We further evaluate the robustness of \textbf{AsarRec} by injecting random noise into user interaction sequences, with noise ratios ranging from 0.1 to 0.5. A noise ratio of 0.1 indicates that 10\% of the original sequence is replaced by randomly sampled irrelevant items, as shown in  Figure~\ref{fig:all_noise}. We find:

\begin{itemize}[leftmargin=*]
    \item \textbf{AsarRec exhibits strong robustness.} AsarRec consistently outperforms all baseline methods under varying noise levels, demonstrating a notably slower performance degradation as noise increases. This confirms its ability to effectively mitigate the impact of noisy interactions.

    \item \textbf{Baseline augmentation methods are highly sensitive to noise.} Static augmentations suffer substantial performance drops as the noise ratio increases. Their lack of adaptivity prevents them from distinguishing between informative and noisy interactions, leading to unstable performance.

    % \item \textbf{Dataset-specific patterns persist under noise.} On \textbf{MIND}, which contains longer sequences and more complex user behaviors, most baseline methods fluctuate sharply across different noise levels. In contrast, AsarRec maintains stable and leading results, underscoring its ability to generalize across datasets with varying complexity.
\end{itemize}

These results confirm that AsarRec not only improves recommendation accuracy in clean settings but also exhibits strong robustness under noisy environments, which is critical for real-world deployment where user feedback is inherently imperfect.

\begin{table}[t]
    \centering
    \caption{Recommendation performance of different argumentation methods on GRU4rec and FLMPrec.}
    \resizebox{0.45\textwidth}{!}{

\begin{tabular}{clcccc}
    \toprule
    \multicolumn{1}{c}{\multirow{2}{*}{\textbf{Noise Setting}}} & \multicolumn{1}{c}{\multirow{2}{*}{\textbf{Model}}}
    
    & \multicolumn{2}{c}{\textbf{HR} } & \multicolumn{2}{c}{\textbf{NDCG} }\\
    \cmidrule(lr){3-4} \cmidrule(lr){5-6} 
    & & \textbf{@10} & \textbf{@20} & \textbf{@10} & \textbf{@20} \\
    \midrule
    \multirow{7}{*}{\textbf{Clean}} &\textbf{GRU4rec}& 0.0345 & 0.0591 & 0.0169 & 0.0231 \\
    &~+\textbf{Crop}& 0.0349 & 0.0604 & 0.0172 & 0.0238 \\
    &~+\textbf{Mask}&\underline{0.0353} &\underline{0.0613} &\underline{0.0178} &\underline{0.0244} \\
    \cmidrule{3-6}
    &~+\textbf{CoSeRe}& 0.0350 & 0.0600 & 0.0177 & 0.0243 \\
    &~+\textbf{S3Rec}& 0.0339 & 0.0595 & 0.0170 & 0.0232 \\
    \cmidrule{3-6}
    &~+\textbf{AsarRec}&\textbf{0.0375**} &\textbf{0.0650**} &\textbf{0.0187*} &\textbf{0.0255**} \\
    &\multicolumn{1}{c}{Gain}& +6.21\% $\uparrow$& +6.07\% $\uparrow$& +4.77\% $\uparrow$& +4.67\% $\uparrow$\\
    \midrule
    \multirow{7}{*}{\textbf{Noisy}} &\textbf{GRU4rec}& 0.0230 & 0.0401 & 0.0115 & 0.0159 \\
    &~+\textbf{Crop}& 0.0228 & 0.0400 & 0.0113 & 0.0156 \\
    &~+\textbf{Mask}& 0.0235 &\underline{0.0411} &\underline{0.0121} & 0.0162 \\
    \cmidrule{3-6}
    &~+\textbf{CoSeRe}& \underline{0.0236} & 0.0409 & 0.0120 &\underline{0.0164} \\
    &~+\textbf{S3Rec}& 0.0229 & 0.0390 & 0.0113 & 0.0155 \\
    \cmidrule{3-6}
    &~+\textbf{AsarRec}&\textbf{0.0259**} &\textbf{0.0446**} &\textbf{0.0132**} &\textbf{0.0179**} \\
    &\multicolumn{1}{c}{Gain}& +9.82\% $\uparrow$& +8.63\% $\uparrow$& +8.75\% $\uparrow$& +8.63\% $\uparrow$\\
     \midrule
    \multirow{7}{*}{\textbf{Clean}} &\textbf{FMLPrec}& 0.0464 & 0.0725 & 0.0243 & 0.0313 \\
&~+\textbf{Crop}&\underline{0.0483} &\underline{0.0803} &\underline{0.0248} &\underline{0.0328} \\
&~+\textbf{Mask}& 0.0438 & 0.0770 & 0.0213 & 0.0294 \\

\cmidrule{3-6}
&~+\textbf{CoSeRe}& 0.0389 & 0.0668 & 0.0194 & 0.0264 \\
&~+\textbf{S3Rec}& 0.0241 & 0.0423 & 0.0115 & 0.0161 \\
\cmidrule{3-6}
&~+\textbf{AsarRec}&\textbf{0.0534**} &\textbf{0.0858**} &\textbf{0.0284**} &\textbf{0.0364**} \\
&\multicolumn{1}{c}{Gain}& +10.56\% $\uparrow$& +6.79\% $\uparrow$& +14.25\% $\uparrow$& +11.15\% $\uparrow$\\
\midrule
\multirow{7}{*}{\textbf{Noisy}} &\textbf{FMLPrec}&\underline{0.0293} & 0.0489 &\underline{0.0149} &\underline{0.0196} \\
&~+\textbf{Crop}& 0.0277 & 0.0472 & 0.0132 & 0.0181 \\
&~+\textbf{Mask}& 0.0263 &\underline{0.0510} & 0.0115 & 0.0176 \\
\cmidrule{3-6}
&~+\textbf{Sub}& 0.0200 & 0.0377 & 0.0101 & 0.0144 \\
&~+\textbf{S3Rec}& 0.0178 & 0.0354 & 0.0079 & 0.0123 \\
\cmidrule{3-6}
&~+\textbf{AsarRec}&\textbf{0.0314**} &\textbf{0.0544**} &\textbf{0.0158**} &\textbf{0.0215**} \\
&\multicolumn{1}{c}{Gain}& +7.15\% $\uparrow$& +6.64\% $\uparrow$& +6.22\% $\uparrow$& +9.53\% $\uparrow$\\

    \bottomrule
\end{tabular}
    }
\label{tab:other_performance}%
\end{table}

\subsection{Argumentation Study~(RQ2)}
\label{sec:rq2}
% In this section, we address \textbf{RQ2} by evaluating the generalization of our method with the other backbone models, visualizing the transformation matrix of our method, as well as user case for study.

% In this section, we address \textbf{RQ2} by evaluating the generalization ability of \textbf{AsarRec} across different backbone models, and conducting a user case study. In addition, we also compare the time complexity of different methods, visualize the AsarRec learned transformation matrices, and ablation study; due to space constraints, the corresponding results are provided in Appendix~\ref{sec:ap_time}.

In this section, we address \textbf{RQ2} by evaluating the generalization ability of \textbf{AsarRec} across different backbone models and conducting a user case study. We further analyze its \textbf{learned transformation matrices}, and \textbf{time complexity}. 

% Due to space constraints, detailed results are provided in Appendix~\ref{sec:ap_time}.

\subsubsection{Generalization Across Various Backbone Models}
We conduct experiments on two additional representative backbones: \textbf{GRU4rec}~\cite{hidasi2015session}, and \textbf{FMLPrec}~\cite{zhou2022filterenhanced}. Due to space constraints, we report results on a subset of five representative augmentation methods.
Table~\ref{tab:other_performance} summarizes the results under both clean (noise ratio = 0.0) and moderately noisy (noise ratio = 0.2) settings. We observe:

\begin{itemize}[leftmargin=*]
    \item \textbf{AsarRec achieves consistent and substantial gains.}

    \item \textbf{Compared augmentation baselines offer limited or inconsistent benefits.} While Mask and Crop occasionally improve over base models, these improvements are marginal and not stable across backbones or noise settings.

    \item \textbf{Strong backbones reduce the utility of naive augmentations.} On FMLPrec, which already incorporates denoising mechanisms by design, most baseline augmentation methods fail to improve performance, especially under the noisy setting. In fact, when the noise ratio is 0.2, the majority of augmentation baselines (e.g., Mask, Sub, S3Rec) perform worse than the FMLPrec.
\end{itemize}

These results confirm that AsarRec not only performs well on the base SASRec model but also generalizes effectively to diverse backbone designs, maintaining its advantage in both clean and noisy environments.

\subsubsection{Transformation Matrix Visualization}

\begin{figure}[t]
    \centering
    \includegraphics[width=0.475\textwidth]{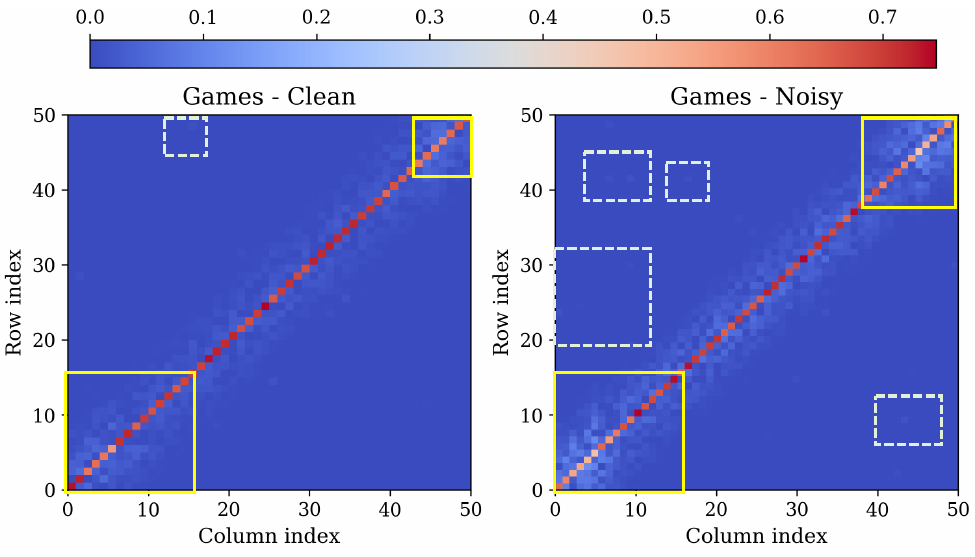}
    \caption{Visualization of the averaged transformation matrix. The yellow solid box highlights strong perturbations near the sequence head and tail, while the gray-green dashed box shows avoidance of long-range reorderings.}
    \label{fig:matrix}
\end{figure}

To better understand the behavior of learned augmentations, we visualize the average transformation matrices of users whose interaction sequence lengths are exactly 50. As shown in Figure~\ref{fig:matrix}:

\begin{itemize}[leftmargin=*]
    \item \textbf{Intensive perturbation at sequence boundaries.} As highlighted by the yellow solid box, stronger transformation activities are observed at the head and tail of sequences, especially near the head. This is consistent with the semantic invariance constraint (Section~\ref{sec:semantic}), as perturbations at the beginning of the sequence typically have less semantic impact on the recommendation task.

    \item \textbf{Increased perturbation under noisy settings.} In high-noise environments, the transformation matrices display more intense and concentrated changes, particularly at the sequence boundaries. This reflects the model's adaptive response to noise—allocating more transformation effort to segments where noise is likely to accumulate while still preserving semantic consistency.

    \item \textbf{Avoidance of long-range reorderings.} As emphasized by the gray-green dashed box, AsarRec tends not to perform long-distance item reordering. This behavior likely arises from the difficulty in maintaining semantic consistency when distant items are swapped, and from the informativeness constraint, which favors localized but challenging perturbations.
\end{itemize}

\begin{figure}[t]
    \centering
    \includegraphics[width=0.475\textwidth]{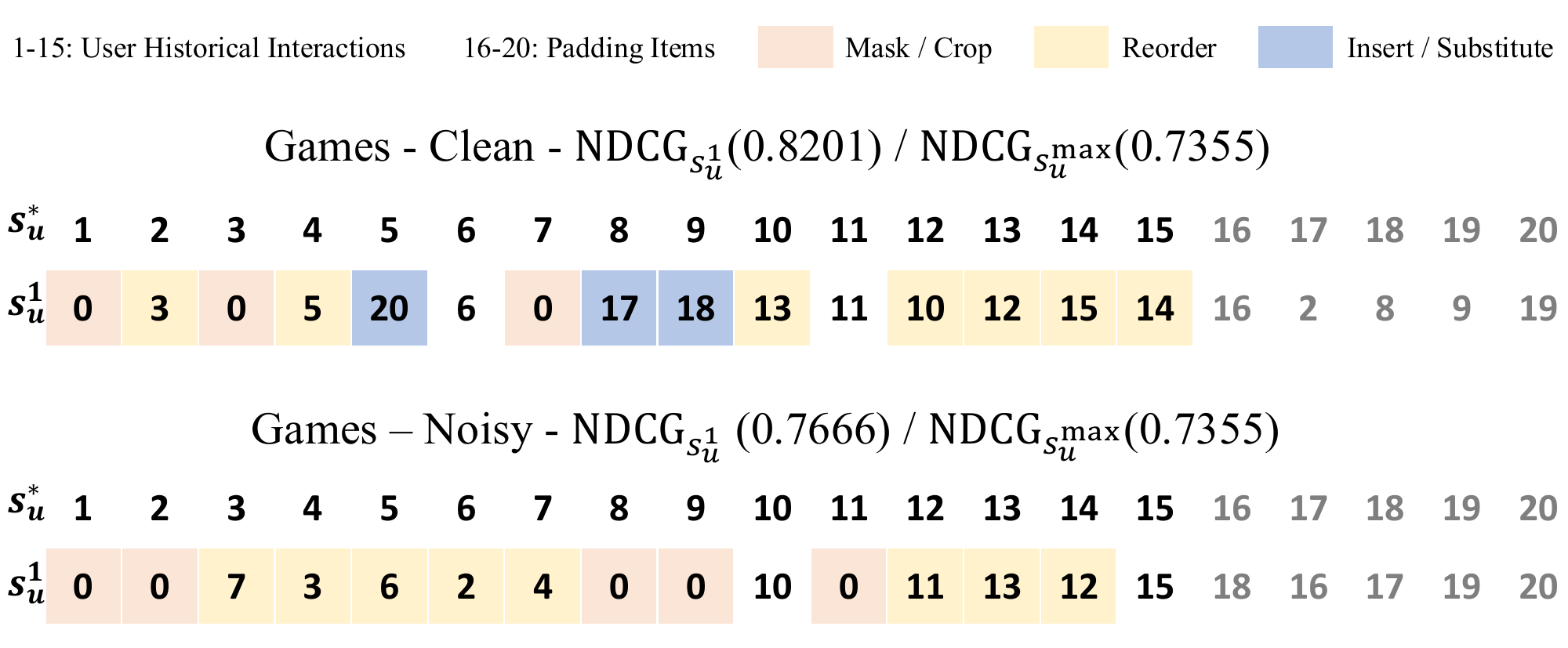}
    \caption{Comparison of the original padded sequence $s_u^*$ and the transformed sequence $s_u'$ for a sampled user from the \textbf{Games} dataset in clean and noisy (20\% noise) settings.}
    \label{fig:user}
\end{figure}

\subsubsection{Case Study}
To gain deeper insights into the behavior of AsarRec under different noise conditions, we conduct a case study on a selected user from the \textbf{Games} dataset. We compare the original padded sequence $s_u^*$ with its transformed sequence $s_u'$ at the 100th training epoch, both in the noise-free setting and under 20\% injected noise. For this user, given a perturbation budget of 10\%, the worst-case semantic degradation threshold $\text{NDCG}^*$ is 0.7355.
As shown in Figure~\ref{fig:user}, several observations can be made:

\begin{itemize}[leftmargin=*]
    \item \textbf{Clean setting.} The perturbations are relatively small, and the resulting $\text{NDCG}$ (0.8201) is much higher than $\text{NDCG}^*$. In this case, AsarRec tends to introduce new interactions through \textit{Insert} and \textit{Substitute} operations, thereby enriching the sequence with new items while preserving its overall order structure.
    
    \item \textbf{Noisy setting.} When 20\% noise is injected, the sequence contains more anomalous or irrelevant interactions. In this scenario, AsarRec becomes more conservative in introducing new interactions and instead favors \textit{Mask} and \textit{Reorder} operations. This behavior effectively removes or repositions suspicious items, reducing the impact of noise and yielding a $\text{NDCG}$ (0.7666).
\end{itemize}
These results illustrate that AsarRec dynamically adjusts augmentation strategies according to the characteristics of the input, enriching sequences in clean contexts and denoising in noisy contexts.

\subsubsection{Time Complexity}

Since the sequence encoder in \textbf{AsarRec} can be pre-trained or trained in parallel, it does not affect the overall training pipeline of the recommender system. Therefore, we analyze the time complexity only during the self-supervised learning stage.

Let $B$ denote the batch size, $L$ the sequence length, and $d$ the hidden dimension.  
For a Transformer-based backbone (e.g., SASRec), the time complexity of one forward pass is $\mathcal{O}(B L^2 d)$, while for an RNN-based backbone (e.g., GRU4Rec) it is $\mathcal{O}(B L d^2)$.

The basic augmentation operations (\textbf{Mask}, \textbf{Crop}, \textbf{Reorder}, \textbf{Insert}, \textbf{Substitute}) are linear with respect to sequence length, i.e., $\mathcal{O}(B L)$, and are thus negligible compared to the backbone’s $\mathcal{O}(B L^2 d)$ cost.

For \textbf{AsarRec}, the additional computations stem from generating probabilistic transition matrices and performing the Semi-Sinkhorn projection, both of which have a complexity of $\mathcal{O}(B L^2)$ with small constants. Consequently, the overall asymptotic complexity of AsarRec remains $\mathcal{O}(B L^2 d)$—the same order as the underlying backbone.

\vspace{0.5em}
\noindent\textbf{Empirical Runtime.}  
Table~\ref{tab:time} reports the average training time per epoch (in seconds) under identical hardware and batch size settings.

\begin{table}[h]
\centering
\caption{Average training time per epoch (mean $\pm$ std) under different augmentation strategies.}
\label{tab:time}
\resizebox{0.40\textwidth}{!}{
\begin{tabular}{lcc}
\toprule
\textbf{Method} & \textbf{Time (s/epoch)} & \textbf{Remarks} \\
\midrule
Backbone & $23.25 \pm 0.67$ & --- \\
Mask & $32.59 \pm 0.96$ & Basic aug. \\
Crop & $33.63 \pm 0.83$ & Basic aug. \\
Reorder & $39.35 \pm 1.37$ & Basic aug. \\
Insert & $34.34 \pm 0.62$ & Basic aug. \\
Substitute & $32.10 \pm 0.37$ & Basic aug. \\
\textbf{AsarRec} & $33.65 \pm 0.69$ & Adaptive aug. \\
\bottomrule
\end{tabular}
}
\end{table}

\vspace{0.5em}
\noindent
Based on the results above, \textbf{AsarRec} does not introduce a significant computational overhead and is even faster than several basic augmentation strategies. Specifically:

\begin{itemize}[leftmargin=*]
    \item Each augmentation must process every user in a batch individually, and since user sequence lengths vary, even basic augmentations incur non-negligible time costs.
    \item \textbf{AsarRec} benefits from efficient tensorized implementation; although it introduces matrix operations, its runtime remains comparable to or faster than several handcrafted augmentations.  
    \item The dominant training cost lies in the computation of self-supervised loss and backward propagation, which are shared by all methods.
\end{itemize}

\subsection{Hyperparameters Analysis \& Ablation~(RQ3)}

% \begin{figure*}
%     \centering
%     \includegraphics[width=0.975\linewidth]{Imgs/hyper.pdf}
%     \caption{Hyperparameter analysis on Games with SASRec backbone: (a) Candidate pool size $k$; (b) Diversity weight $\lambda_{\text{div}}$; (c) Semantic invariance weight $\lambda_{\text{NDCG}}$.}
%     \label{fig:hyper}
% \end{figure*}

\begin{figure}[t]
    \centering
    \includegraphics[width=\linewidth]{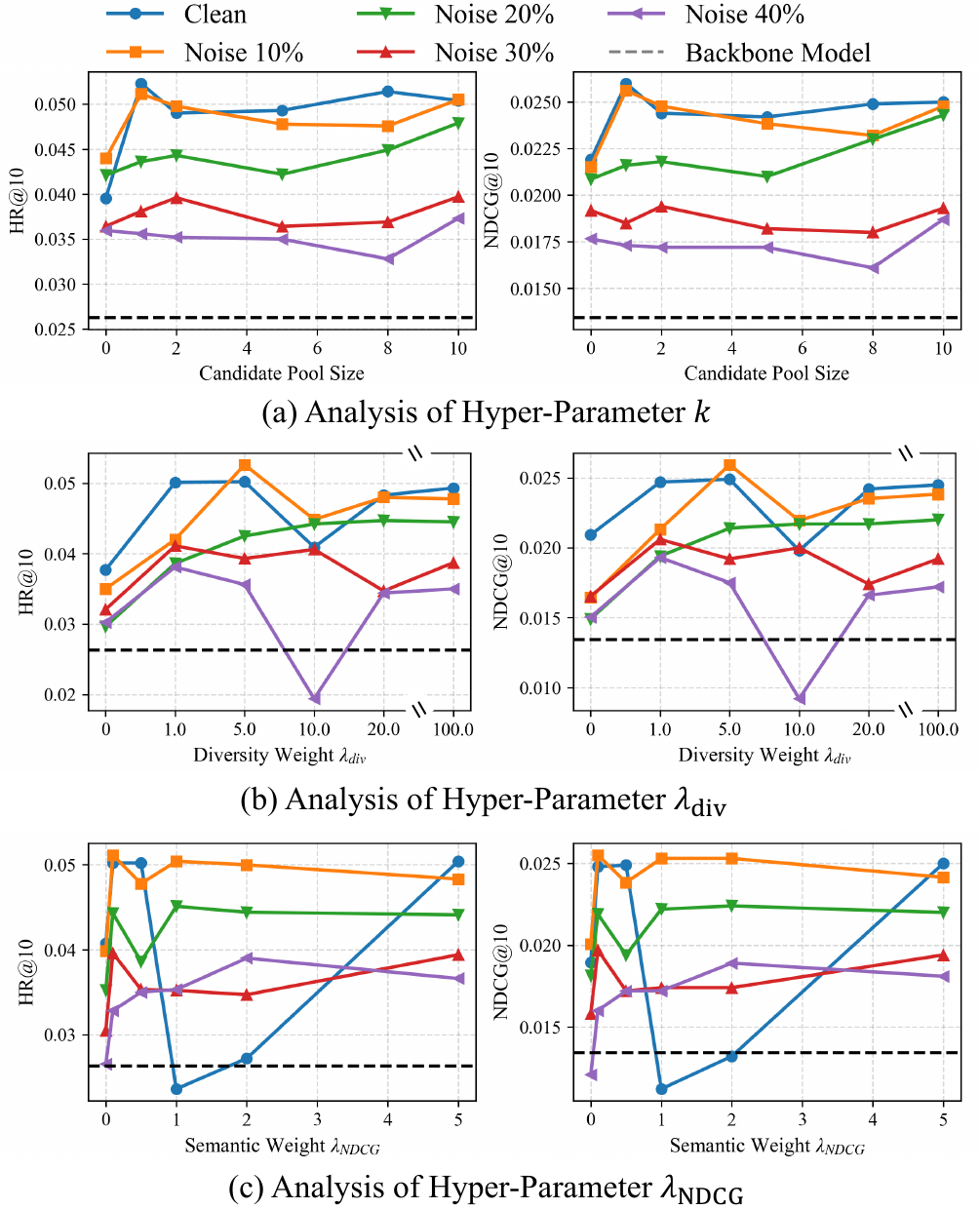}
    \caption{Hyperparameter analysis and ablation study.}
    \label{fig:hyper}
\end{figure}

In this section, we address \textbf{RQ3} by examining the impact of key hyperparameters on the Games dataset using SASRec as the backbone model in Figure~\ref{fig:hyper}. Note that when the horizontal axis equals $0$, it indicates the complete removal of the corresponding constraint, representing an \textbf{ablation} setting.

\subsubsection{Effect of Candidate Pool Size $k$}
The parameter $k$ controls the number of candidate positions considered when constructing the transformation matrix. As shown in Figure~\ref{fig:hyper}(a), model performance remains largely stable across different $k$ values, indicating that \textbf{AsarRec} is insensitive to the candidate pool size. Larger $k$ offers slightly more flexibility, but even small candidate sets yield comparable results, reflecting the robustness of AsarRec.

\subsubsection{Effect of Diversity Weight $\lambda_{\text{div}}$}
The parameter $\lambda_{\text{div}}$ encourages the two augmented views to differ from each other. As shown in Figure~\ref{fig:hyper}(a), model performance remains largely stable across different $k$ values, indicating that \textbf{AsarRec} is insensitive to the candidate pool size. Larger $k$ offers slightly more flexibility, but even small candidate sets yield comparable results, reflecting the robustness of the learned transformations.

\subsubsection{Effect of Semantic Invariance Weight $\lambda_{\text{NDCG}}$}
The parameter $\lambda_{\text{NDCG}}$ controls how well the transformation preserves the original ranking semantics. As illustrated in Figure~\ref{fig:hyper}(c), small $\lambda_{\text{NDCG}}$ values lead to excessive semantic distortion between augmented views, causing notable performance drops—especially under clean settings where preserving sequence order is crucial. When $\lambda_{\text{NDCG}} > 5$, semantic consistency is effectively maintained, and performance becomes stable across all metrics.

\textbf{Ablation.} As shown in Figure~\ref{fig:hyper}, removing the candidate pool size $k$ in low-noise settings leads to a substantial performance drop, indicating its importance for providing sufficient transformation flexibility. Similarly, omitting either $\lambda_{\text{div}}$ or $\lambda_{\text{NDCG}}$ results in noticeable degradation in overall performance. These results confirm that each component of \textbf{AsarRec} plays a critical and complementary role in achieving robust augmentations.

\section{CONCLUSION}
% In this work, we present \textbf{AsarRec}, a novel adaptive augmentation framework for robust self-supervised sequential recommendation. Motivated by the limitations of existing static augmentation methods---which fail to generalize across datasets and noise settings---we propose to unify sequence augmentation operations as structured transformation matrices. Based on this formulation, AsarRec learns personalized augmentation strategies by generating transformation matrices through a differentiable semi-Sinkhorn process. To ensure the effectiveness of the learned augmentations, we further introduce three principled optimization constraints: \textit{diversity}, \textit{semantic invariance}, and \textit{informativeness}. 
% Extensive experiments on multiple real-world datasets under various noise conditions demonstrate that AsarRec consistently outperforms existing augmentation baselines and state-of-the-art methods. Our results validate that AsarRec can dynamically adapt augmentation behavior to both user-specific interaction patterns and dataset characteristics, making it a general and robust solution for self-supervised sequential recommendation. 

In this work, we propose \textbf{AsarRec}, an adaptive augmentation framework for robust self-supervised sequential recommendation. To address the limitations of existing static augmentation methods that fail to generalize across datasets and noise levels, we unify sequence augmentation operations into a structured matrix transformation formulation. Building on this, AsarRec learns personalized augmentation strategies via a differentiable semi-Sinkhorn process and optimizes them through three key constraints: \textit{diversity}, \textit{semantic invariance}, and \textit{informativeness}. Extensive experiments on multiple real-world datasets demonstrate that AsarRec consistently outperforms existing baselines, effectively adapting to user-specific patterns and dataset characteristics to achieve robust and generalizable recommendation performance.

\bibliographystyle{ACM-Reference-Format}
\bibliography{ref}

% \clearpage
% \appendix
% \section{APPENDIX}
% \input{Section/Appendix}

\end{document}